\pdfoutput=1
\documentclass[11pt,a4paper]{article}
\usepackage[T1]{fontenc}
\usepackage[utf8]{inputenc}
\usepackage[margin=2.5cm]{geometry}
\usepackage{amsmath,amssymb,amsthm}
\usepackage{graphicx}
\usepackage{booktabs,tabularx}
\usepackage{enumitem}
\usepackage{float}
\usepackage[numbers,sort&compress]{natbib}
\usepackage[colorlinks=true,allcolors=blue]{hyperref}

\setlist[itemize]{leftmargin=*,labelsep=5.8mm}
\newtheorem{Proposition}{Proposition}
\newtheorem{Lemma}{Lemma}

\newcommand{\nabs}{n_{\mathrm{abs}}}

\title{A loss-limited optimum for fixed uniform multi-pass quantum sensing per absorbed photon}
\author{Christoph F. Wildfeuer\thanks{Correspondence: christoph.wildfeuer@fhnw.ch}\\[4pt]
\parbox{0.9\textwidth}{\centering\normalsize Institute for Sensors and Electronics, School of Engineering, FHNW University of Applied Sciences and Arts Northwestern Switzerland, Klosterzelgstrasse 2, CH-5210 Windisch, Switzerland}}
\date{}

\begin{document}
\maketitle

\begin{abstract}
Multi-pass schemes send a photon through a sample several times to learn more about it. When the sample rather than the light is scarce, the natural figure of merit is the information gained per photon the object absorbs. We show that three properties fix the loss-limited optimum of every fixed scheme in which a single photon passes repeatedly and uniformly through the sample and is detected once at the end: a payoff that grows as the square of the pass number, the same survival probability on every pass, and no dose from a photon already lost. They force a single trade-off function $h(x)=x^{2}/(e^{x}-1)$, where $x$ is the number of passes times the loss per pass. Its maximum, $0.648$, sits at $x_{\rm opt}=1.594$. The multi-pass optima of Koppell and Kasevich and of Yu \emph{et al.} are two instances, and the loss-limited ceiling of interaction-free interrogation is a third when its payoff is counted as conclusive detections. Phase sensing of a weakly absorbing object is optimal at $m_{\rm opt}=x_{\rm opt}/(\epsilon+\alpha)$ passes; the absorption $\alpha$ and the parasitic loss $\epsilon$ enter the optimum only through their sum, but the damage counts only $\alpha$. N00N states and their loss-robust generalisations do worse per absorbed photon, and optimising over photon number and pass number together returns a single recycled photon; Gaussian probes and chains tailored pass by pass lie outside this class and can exceed its ceiling by a bounded factor. Two further problems are priced in the same measure. Absorption estimation gains nothing, for any scheme. Above the crosstalk crossover of the mode sorter, detecting a faint companion below the Rayleigh limit costs the companion a dose that does not depend on its faintness; below it, and under direct imaging at any separation, the dose grows without bound as the companion fades. Among single photons and path-entangled number states, the gentlest measurement of a fragile sample is also the simplest one: a single photon, recycled.
\end{abstract}


\section{Introduction}
\label{sec:intro}

Multi-pass schemes send a probe through an object more than once, and each additional pass buys signal while it risks the photon, so there is a best number of passes and the loss sets it. Such optima have been computed for several problems, in electron microscopy, in optical interferometry and in interaction-free measurement. They are the same number.

Throughout, $\alpha$ is the probability that the object absorbs the photon on one pass, $\epsilon$ the parasitic loss per pass outside the object, $L=\epsilon+\alpha$ the total loss per pass and $\eta=q=1-L$ the survival per pass; $m$ or $N$ counts the passes and $x=-m\ln q\simeq mL$ is the number of passes times the loss per pass. For dose-limited phase estimation, with electron microscopy in view, Koppell and Kasevich showed that a fixed multi-pass chain reaches $0.65$ of the quantum limit on information per unit dose at the optimal pass number, and that a N00N state reaches $1/e$ of it \cite{Koppell2022,KoppellKasevich2022}. Dose-efficient multi-pass phase interferometry attains its optimum at $N_{\rm opt}\simeq 1.6/\!\ln(1/\eta)$ passes, with an optimality coefficient $0.648$ \cite{Yu2026}. The interaction-free interrogator of Elitzur, Vaidman and Kwiat delivers conclusive detections per absorbed photon at a rate that parasitic per-cycle loss $\epsilon$ caps at $0.2625/\epsilon$, attained at $N_{\rm opt}=1.5936/\epsilon$ cycles \cite{Kwiat1999,Wildfeuer2026}. Phase sensing of a weakly absorbing object, treated below, is optimal at $m_{\rm opt}=x_{\rm opt}/(\epsilon+\alpha)$ passes with a gain $0.6476/(\epsilon+\alpha)$. Note that $1.5936$ and $1.6$ are the same number, that $0.65$ and $0.648$ are the same number, and that $0.2625$ is $0.648$ divided by $\pi^{2}/4$.

All three reduce to maximising
\begin{equation}
h(x) \;=\; \frac{x^{2}}{e^{x}-1}.
\label{eq:h}
\end{equation}
The maximum sits at the root $x_{\rm opt}=1.59362$ of $2(e^{x}-1)=x\,e^{x}$, where $h(x_{\rm opt})=0.64761$. The Zeno ceiling carries in addition the geometric factor $\pi^{2}/4$ of the chain. In discrimination the constant caps a rate; in phase estimation it multiplies a sensitivity. Since $x$ is the number of passes times the loss per pass, one percent loss per pass puts the optimum at $m_{\rm opt}\approx159$ passes, where the photon reaches its single detection with probability $e^{-x_{\rm opt}}\approx0.20$. Up to that point the quadratic growth of the signal beats the exponential decay of the survival. Beyond it the decay wins.

What makes the three comparable is the choice of resource. Reference \cite{Wildfeuer2026} priced interaction-free measurement in information per photon absorbed by the object. That is the natural measure whenever the sample and not the light is scarce: a photosensitive specimen, a biological structure, a sample that x-rays damage. We show here that the same measure organises fixed uniform multi-pass sensing, and we say exactly which schemes it does not cover.

Section \ref{sec:currency} fixes the resource measure and recalls why the absorption parameter admits no super-linear gain for anyone. Section \ref{sec:universal} states the optimum, then applies it to the published cases and to a new one. In the new case the object absorbs as well as retards, and the two losses play different roles: they enter the optimum only through their sum, while the damage counts only the absorption. Section \ref{sec:subrayleigh} prices a fourth problem in the same measure. Deciding whether a faint companion sits beside a known source, below the Rayleigh limit, is a null experiment in the sense of the lemma. Its ideal exponents and its crosstalk ceiling are known \cite{LuKroviTsang2018,Schlichtholz2024,HuangLupo2021}. Priced per absorbed photon they yield one new invariance: above the crosstalk crossover, what discovery costs the companion does not depend on how faint the companion is, while under direct imaging that cost grows without bound. Sections \ref{sec:prior} to \ref{sec:conclusion} delimit prior work, give the experimental numbers, and collect the rules.

The scope throughout is fixed uniform arrangements: the same coupling on every pass and one detection at the end. We treat single-photon multi-pass chains, and in Sec.~\ref{sec:entangled} path-entangled number states, which do not reach the same optimum. Three kinds of scheme lie outside the class and can beat its ceiling: adaptive control \cite{Yu2026}, chains whose coupling is tailored pass by pass, and Gaussian probes such as displaced squeezed states \cite{Koppell2022,KoppellKasevich2022}. We say in Sec.~\ref{sec:universal} which hypothesis each violates.

\section{Information per absorbed photon}
\label{sec:currency}

\subsection{The resource measure}

All figures of merit below are normalised by $\nabs$, the mean number of photons absorbed by the object under study; photons sent, photons collected, and photons lost elsewhere in the apparatus do not enter. For a fragile sample $\nabs$ is the number that ends the experiment. It is not the ``dose'' of Ref.~\cite{Yu2026}, which counts every photon entering the sample channel. The two agree only when the sample dominates the loss budget, and they differ precisely in the regime minimally invasive sensing is meant for --- a nearly transparent object behind lossy optics. The measure is not tied to single photons. It applies to any probe state, and Secs.~\ref{sec:absorption} and \ref{sec:entangled} use it for arbitrary and for entangled states. Information is measured by the Fisher information for estimation \cite{Fisher1925,BraunsteinCaves1994} and by the Chernoff information for discrimination \cite{Chernoff1952,CoverThomas2006}; the quantum Fisher information is the Fisher information optimised over all measurements of the state \cite{BraunsteinCaves1994}.

\subsection{Absorption admits no super-linear gain}
\label{sec:absorption}

Before turning to phase, one parameter can be disposed of in this measure for every probe state at once. For the transmissivity $\eta$ of a loss channel no strategy achieves super-linear scaling, entangled, adaptive or counterfactual. The quantum Fisher information is linear in the mean photon number sent into the channel, $F_Q \le \bar n/[\eta(1-\eta)]$, and Fock states saturate it \cite{MonrasParis2007,Adesso2009,Nair2018,WangDavidovichAgarwal2020}. That bound is per photon sent. Per absorbed photon it reads $F_Q/\nabs\le1/[\eta(1-\eta)^{2}]$, since $\nabs=\bar n(1-\eta)$ for a single pass; the two measures differ by a factor that does not depend on the probe, so the conclusion is the same in both. The fixed sequential single-photon case is closed by the theorem of Ref.~\cite{Wildfeuer2026}. Supersensitive counterfactual estimation of transparency is therefore excluded, not merely unobserved. Everything that follows lives in the two places this no-go leaves open: phase, where coherence accumulates (Sec.~\ref{sec:universal}), and discrimination, where a null experiment applies (Sec.~\ref{sec:subrayleigh}).

\section{The loss-limited optimum}
\label{sec:universal}

\subsection{Why $h(x)$ is forced}

\begin{Proposition}
Let a scheme in the fixed uniform single-photon multi-pass class satisfy, for pass numbers $m\ge1$ and per-pass survival $q\in(0,1)$, $L=1-q$:
\begin{itemize}
\item[(H1)] the payoff per incident photon is $A\,m^{2}q^{m}$, with $A>0$ independent of $m$ and $q$, the payoff being the Fisher information for estimation and the conclusive-detection probability for null discrimination;
\item[(H2)] per-pass survival is $q$, the same on every pass and independent across passes, so the arrangement is uniform and the object is memoryless;
\item[(H3)] the resource accumulates only along surviving paths, $\mathrm{cost}=B\,(1-q^{m})/(1-q)$ with $B>0$.
\end{itemize}
Then, with $x=-m\ln q$,
\begin{equation}
\frac{\mathrm{payoff}}{\mathrm{cost}} \;=\; \frac{A}{B}\cdot\frac{1-q}{\ln^{2}q}\,h(x),
\qquad h(x)=\frac{x^{2}}{e^{x}-1},
\label{eq:prop}
\end{equation}
exactly.
\end{Proposition}
Over continuous $m$ the maximum sits at $m_{\rm opt}=x_{\rm opt}/(-\ln q)$ with value $(A/B)\,h(x_{\rm opt})\,(1-q)/\ln^{2}q$, where $x_{\rm opt}=1.59362$ solves $2(e^{x}-1)=xe^{x}$. For $L\ll1$ the prefactor is $(1-q)/\ln^{2}q=1/L+\mathcal{O}(1)$, so $m_{\rm opt}=x_{\rm opt}/L$ and the value is $(A/B)h(x_{\rm opt})/L$. The pass number is a positive integer, and $h$ is unimodal, so the maximum over $m\in\{1,2,\dots\}$ sits at $\lfloor m_{\rm opt}\rfloor$ or $\lceil m_{\rm opt}\rceil$ if $m_{\rm opt}\ge1$, and at $m=1$ if $m_{\rm opt}<1$, that is if $q<e^{-x_{\rm opt}}=0.203$. Comparing $m=1$ with $m=2$ sharpens the boundary: a single pass is optimal if and only if $q\le1/3$, so multi-passing pays only for a per-pass loss $L<2/3$. For $m_{\rm opt}\ge1$ the integer restriction lowers the optimised value by a relative amount of at most $0.058\,\ln^{2}q$ to leading order in $\ln q$, the coefficient being $\tfrac18\,|(\ln h)''(x_{\rm opt})|$. Numerically the reduction stays below $6.2\times10^{-4}$ for all $L\le0.1$ and below $0.74\%$ for all $L\le0.3$, and it never exceeds $6.8\%$, the value at the tie $q=1/3$. \begin{proof} Write $m^{2}q^{m}=x^{2}e^{-x}/\ln^{2}q$ and $1-q^{m}=1-e^{-x}$. The sign of $h'(x)$ is that of $\phi(x)=2(e^{x}-1)-xe^{x}$. Since $\phi(0)=0$ and $\phi'(x)=e^{x}(1-x)$, $\phi$ is positive on $(0,x_{\rm opt})$ and negative beyond, so $h$ is unimodal. For the boundary, the ratio of the payoff per cost at $m=2$ to that at $m=1$ is $4q/(1+q)$, which exceeds one if and only if $q>1/3$. \end{proof}

The proposition is a substitution, and its content is not the algebra. It is the three hypotheses, which are the structure that Refs.~\cite{KoppellKasevich2022,Yu2026,Wildfeuer2026} share without saying so, and which make the result predictive: any fixed uniform scheme satisfying (H1) to (H3) has its optimum at $x_{\rm opt}$ and its ceiling at $h(x_{\rm opt})(1-q)/\ln^{2}q$ times its single-pass figure, whatever the implementation. The hypotheses hold exactly for coherent phase accumulation in the model of Sec.~\ref{sec:phase}. The Zeno chain satisfies them only asymptotically, for $N\to\infty$ at fixed $N\epsilon$, because its survival and its absorption probability per cycle both depend on the cycle number (Appendix~\ref{app:zeno}). Physically the quadratic law arises whenever the signal is a coherent sum over passes --- a phase, a displacement, a pointer shift --- so that amplitudes rather than probabilities add. The Zeno chain reaches the same algebra by a different route: its conclusive probability is $q^{m}$ alone, but the absorption probability per cycle falls as the inverse square of the cycle number, so the factor $m^{2}$ sits in the dose rather than in the payoff. Only the ratio $A/B$ enters Eq.~(\ref{eq:prop}), and the proposition gives the leading order of the chain's rate in $\epsilon$ (Appendix~\ref{app:zeno}). The distinction is kept throughout: statements under (H1) to (H3) are exact in $x=-m\ln q$, and statements about a particular implementation are exact only where we say so, and otherwise hold to leading order in the small parameter named.

Table~\ref{tab:classes} locates the known cases. A linear payoff with independent Bernoulli encounters, which is absorption estimation, gives a ratio independent of $m$: the flat exchange rate of Ref.~\cite{Wildfeuer2026}. A linear payoff with conclusive-outcome counting, which is the Zeno chain, grows logarithmically and is capped by loss through $h$. A quadratic payoff, which is coherent phase accumulation, is the case above. Outside the class the hypotheses fail one at a time, and identifiably. Adaptive control breaks the fixed-arrangement premise behind (H1) and can beat the ceiling \cite{Yu2026}. A chain whose coupling is tailored pass by pass, such as the chain interferometer of Ref.~\cite{KoppellKasevich2022}, whose beam-splitter transmissivities fall geometrically along the chain, $T_k\propto\eta^{\,m-k}$, breaks the uniformity in (H2) and approaches the quantum limit for large $m$. A Gaussian probe breaks (H1), since its information is not a single-photon $m^{2}q^{m}$; Refs.~\cite{Koppell2022,KoppellKasevich2022} show that squeezed Gaussian probes saturate the per-dose quantum limit in the limit of strong squeezing, a limit of which the uniform single-photon chain reaches the fraction $0.65=h(x_{\rm opt})$. An object with memory breaks (H2), and post-selected dose accounting breaks (H3). The ceiling of this paper is therefore the ceiling of its class. Gaussian and tailored probes can exceed it, by at most the factor $1/h(x_{\rm opt})\approx1.54$ that separates the class from the quantum limit of Refs.~\cite{Koppell2022,KoppellKasevich2022}.

\begin{table}[H]
\caption{\label{tab:classes}The four cases, priced per absorbed photon. The dose column decides the constant: weighting by survival gives $x_{\rm opt}=1.594$, and no weighting gives $x_{\rm opt}=1$. The absorption row has no optimum at all; its ratio is a flat exchange rate. The optima are leading-order values for small loss: Eq.~(\ref{eq:prop}) is exact for the phase chain, and the Zeno row is the leading order of an asymptotic relation (Appendix~\ref{app:zeno}).}
\begin{tabularx}{\textwidth}{llll}
\toprule
payoff & dose & optimum & realisation \\
\midrule
linear, Bernoulli & weighted & none & absorption \cite{Wildfeuer2026} \\
linear, conclusive & weighted & $0.2625/\epsilon$ & Zeno chain \cite{Wildfeuer2026} \\
quadratic, sequential & weighted & $0.6476/(\epsilon{+}\alpha)$ & multi-pass phase \cite{KoppellKasevich2022,Yu2026} \\
quadratic, parallel & unweighted & $0.3679/(\epsilon{+}\alpha)$ & N00N \cite{Lee2009,KoppellKasevich2022,Yu2026} \\
\bottomrule
\end{tabularx}
\end{table}

\subsection{Interaction-free discrimination}
\label{sec:zenosec}

The Zeno interrogator of Refs.~\cite{Kwiat1999,Wildfeuer2026} delivers conclusive interrogations per absorbed photon at the lossless rate $4N/\pi^{2}$. Per-cycle loss $\epsilon$ caps it, to leading order in $\epsilon$, at
\begin{equation}
N_{\rm opt}\simeq\frac{x_{\rm opt}}{\epsilon},\qquad
\mathcal{R}_{\max}\simeq\frac{4}{\pi^{2}}\,\frac{h(x_{\rm opt})}{\epsilon}=\frac{0.2625}{\epsilon},
\label{eq:zeno}
\end{equation}
which is Eq.~(17) of Ref.~\cite{Wildfeuer2026}. The chain's exact rate is $\mathcal{R}_{\epsilon}(N)=u^{N}(1-u)/[\sin^{2}(\pi/2N)(1-u^{N})]$ with $u=(1-\epsilon)\cos^{2}(\pi/2N)$. It has the algebraic form of the proposition with $A/B=4/\pi^{2}$, but the chain does not satisfy the hypotheses in their exact form, since the survival $u$ and the absorption probability $\sin^{2}(\pi/2N)$ both depend on $N$. The relation is asymptotic. For $N\to\infty$ at fixed $N\epsilon$ the rate is $(4/\pi^{2})h(N\epsilon)/\epsilon$ up to a relative correction of order $1/N$, and at the optimum $\mathcal{R}_{\max}=(0.2625/\epsilon)\,[1-1.97\,\epsilon+\mathcal{O}(\epsilon^{2})]$; Appendix~\ref{app:zeno} derives both. The quadratic factor arises here for a different physical reason than in the phase case: not coherent accumulation, but an absorption probability per cycle, $\sin^{2}(\pi/2N)\simeq\pi^{2}/4N^{2}$, that falls as the inverse square of the cycle number. Two distinct mechanisms reach the same algebraic class, which is why one constant governs both.

The payoff counted here is the conclusive-detection rate, as in Ref.~\cite{Wildfeuer2026}, not an information. The two are related by the lemma of Sec.~\ref{sec:null}: a trial that is conclusive with probability $p$ carries $-\ln(1-p)\ge p$ nats of Chernoff information, so the information per absorbed photon exceeds the rate, and its own maximum sits elsewhere, at $x=1.270$ with value $0.739$, that is $0.300/\epsilon$ after the factor $4/\pi^{2}$. The rate optimum $x_{\rm opt}$ delivers $98\%$ of that. The identity of the constants across Table~\ref{tab:classes} is thus asymptotically exact, for $\epsilon\to0$, for the conclusive-detection rate, and holds to two percent for the Chernoff information.

\subsection{Phase estimation on a weakly absorbing object}
\label{sec:phase}

Let the object retard by $\varphi$ and absorb with probability $\alpha\ll1$ per pass, and let the photon be folded $m$ times through it by a mirror loop, with parasitic per-pass loss $\epsilon$ and a single detection after the last pass. The passes are one unitary evolution; nothing is measured in between. A photon the object absorbs is gone, and the bookkeeping below counts exactly those. The phase accumulates coherently. Let the photon enter the sensing arm with probability $w$ and the reference arm with $1-w$, let the sensing arm traverse the object $m$ times with per-pass survival $q=(1-\epsilon)(1-\alpha)$, and measure information in units where a single lossless pass carries one unit per photon in the sensing arm. The quantum Fisher information of the lossy single-photon state is then $4m^{2}w(1-w)q^{m}/(1-w+wq^{m})$ per incident photon, attained by a fringe readout at its best working point, while the dose is $w$ times that of a photon sent wholly through the object. The ratio of the two decreases monotonically in $w$, so the optimal readout per absorbed photon is the small-weight limit $w\to0$, in which the information per photon in the sensing arm is $m^{2}q^{m}$. This is the unbalanced limit of Ref.~\cite{Yu2026}, and it is the convention used throughout; the entangled-state comparison of Sec.~\ref{sec:entangled} is quoted in both conventions. A balanced interferometer, $w=\tfrac12$, gives instead $2m^{2}q^{m}/(1+q^{m})$ per incident photon at half the dose. Its trade-off function is $x^{2}/(2\sinh x)$, with optimum $x=1.915$ and value $0.552$, which is $15\%$ below $h(x_{\rm opt})$: the constant $0.648$ belongs to the optimal readout, not to the apparatus. The dose is what the object absorbs, and it accumulates only over the passes the photon survives to make,
\begin{equation}
\nabs = \alpha(1-\epsilon)\,\frac{1-q^{m}}{1-q}.
\label{eq:dose}
\end{equation}
Hypotheses (H1) to (H3) hold exactly here, with $A=1$ and $B=\alpha(1-\epsilon)$, so that $F_\varphi/\nabs=[\alpha(1-\epsilon)]^{-1}\,[(1-q)/\ln^{2}q]\,h(-m\ln q)$ exactly. Writing $L=\epsilon+\alpha$, to leading order in $L$ this is
\begin{equation}
\begin{split}
\frac{F_\varphi}{\nabs} &\simeq \frac{1}{\alpha}\cdot\frac{1}{L}\,h(mL),\\
m_{\rm opt}\simeq\frac{x_{\rm opt}}{\epsilon+\alpha},
&\qquad
G_{\max}\simeq\frac{h(x_{\rm opt})}{\epsilon+\alpha},
\end{split}
\label{eq:gmax}
\end{equation}
where $G_{\max}$ is the gain over single-pass probing at equal dose. Numerical maximisation of the exact expression confirms this to better than $1\%$ for $\epsilon,\alpha\le10^{-2}$.

Two features deserve note. First, the two losses separate. Both remove the photon and enter the survival symmetrically, $q=(1-\epsilon)(1-\alpha)\simeq1-L$; but only $\alpha$ is deposited in the sample, so only $\alpha$ counts as dose. They enter the optimum only through their sum, while the resource measure counts only $\alpha$. The separation is elementary, and we do not claim it as a result; a treatment with a single loss channel simply cannot see it. Second, the bookkeeping in Eq.~(\ref{eq:dose}) matters. Assuming instead that a photon doses on every pass regardless of prior loss, $\nabs\simeq m\alpha$, underestimates the gain by a factor $e\,h(x_{\rm opt})\approx1.76$ and puts the optimum at $1/\epsilon$.

The multi-pass optimum itself was found before, by Koppell and Kasevich with a survival-weighted dose \cite{Koppell2022,KoppellKasevich2022}, and independently by Yu and co-workers for optical interferometry \cite{Yu2026}. The fraction $0.65$ of the quantum limit in the former and the $N_{\rm opt}\simeq-1.6/\ln\eta$ with coefficient $0.648$ in the latter are $h(x_{\rm opt})$ and $x_{\rm opt}$ for a single loss channel $\eta=1-L$. What this section adds is the separation of $\alpha$ from $\epsilon$, the absorbed-photon measure in place of photons entering the sample, and the identification through Eqs.~(\ref{eq:h}) and (\ref{eq:zeno}) that their optimum and the interrogator's are one number.

How pure can a phase object be? The Kramers--Kronig relations link dispersion and absorption only as an integral over frequency. At a single probe frequency $\alpha$ can therefore be made arbitrarily small at fixed $\varphi$ --- window glass is the everyday proof --- and the far-off-resonant limit is the legitimate phase-object regime. For a resonant sample the two are locked by the line shape instead. For a Lorentzian line of width $\Gamma$ probed at detuning $\Delta$, both scale with the number density $n$ of absorbers, $\alpha=\alpha_{0}/(1+D^{2})$ and $\varphi=\alpha_{0}D/(1+D^{2})$ with $D=\Delta/\Gamma$ and $\alpha_{0}\propto n$ the resonant absorption, so $\varphi/\alpha=D$. What is estimated is $n$, and its Fisher information per absorbed photon is $(\partial\varphi/\partial n)^{2}F_{\varphi}/\nabs$. For a single pass this is proportional to $\varphi^{2}/\alpha\propto D^{2}/(1+D^{2})$, which saturates: beyond $\Delta=\Gamma$ detuning buys at most a factor of two. Multi-pass changes this. By Eq.~(\ref{eq:gmax}) the information per absorbed photon is proportional to $[D^{2}/(1+D^{2})]\,h(x_{\rm opt})/(\epsilon+\alpha)$, and the second factor keeps growing as detuning drives $\alpha$ down, until the sample's absorption falls to the parasitic loss. Detuning therefore pays until $\alpha\approx\epsilon$, that is until $D\approx\sqrt{\alpha_{0}/\epsilon}$, and buys at most a further factor of two beyond it, where the gain over single-pass probing approaches the full $h(x_{\rm opt})/\epsilon$ of Eq.~(\ref{eq:gmax}). We confirmed this numerically with the exact expressions: for $\alpha_{0}=0.1$ and $\epsilon=10^{-3}$ the information per absorbed photon at $m_{\rm opt}$ rises by a factor $53$ between $D=1$ and $D=10$ and by only $2$ beyond. In practice the absolute phase per pass also falls as $1/D$, so detector dark counts and the available integration time set a lower limit on the signal that may stop the detuning earlier. The rule for spectroscopy of a fragile sample is therefore to detune until the sample's absorption matches the loss of the optics, and then spend what remains on $m_{\rm opt}$ passes.

A phase probe traverses the sample, so the counterfactual label does not apply here. What carries over from Ref.~\cite{Wildfeuer2026} is the resource measure and the optimum, not any interaction-free claim.

\subsection{Entangled probes spend the resource differently}
\label{sec:entangled}

Does the same constant govern entangled schemes? It does not, and they do worse.

\begin{figure}[H]
\includegraphics[width=0.95\textwidth]{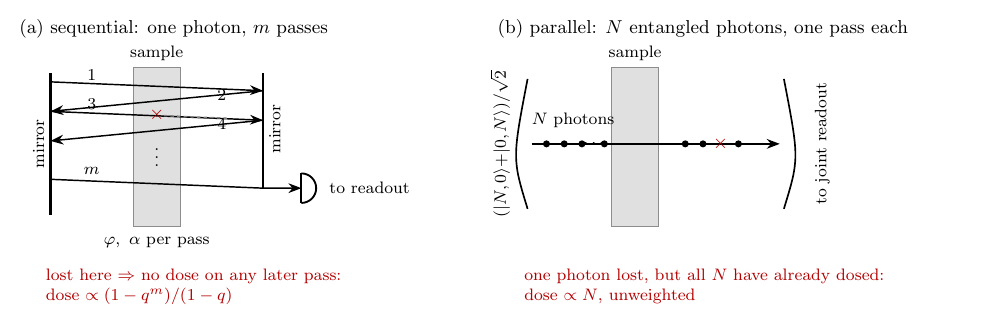}
\caption{\label{fig:schemes}Two ways to spend $Nm$ probe uses on a sample. (a) A single photon folded $m$ times through the sample, detected once. A photon absorbed on one pass makes no later pass, so the dose is weighted by survival. (b) A maximally path-entangled state of $N$ photons, each meeting the sample once. The coherence requires all $N$ photons to survive, but every photon doses the sample whatever becomes of the others. The weighting is hypothesis (H3), and it is the entire difference between the constants $0.648$ and $1/e$. Only the sensing arm is drawn: in both schemes it is one arm of an interferometer, and the reference arm and the recombination that reads the phase are omitted. In (b) the drawn branch is the $|N,0\rangle$ component of the superposition: all $N$ photons share one mode.}
\end{figure}

Take the parallel counterpart of the sequential chain, a maximally path-entangled number state of $N$ photons. The photons do not travel on $N$ separate paths: all $N$ share one mode, and that mode traverses the object once. Parallel here means parallel in time --- the whole budget of probe uses is spent in a single traversal. The phase accumulates as $N\varphi$, so the payoff is again quadratic, $F_\varphi = N^{2}q^{N}$, the factor $q^{N}$ being the decay of the coherence under per-photon survival $q$ \cite{Lee2009}. Hypothesis (H1) holds, and so does (H2). Hypothesis (H3) fails. Each of the $N$ photons deposits its dose in that single traversal, whether or not the state's coherence survives, so the dose is proportional to $N\alpha$ and carries no survival weighting. In a sequential chain a photon lost at the third pass cannot deposit anything on the seventh; in the one-shot state there is no such protection (Fig.~\ref{fig:schemes}).

The optimisation is then over $N q^{N}$ rather than over $h(NL)$, and it gives, to leading order in $L$,
\begin{equation}
N_{\rm opt}\simeq\frac{1}{L},
\qquad
G^{\mathrm{par}}_{\max}\simeq\frac{1}{e\,L}=\frac{0.3679}{L},
\label{eq:parallel}
\end{equation}
which exhaustive maximisation confirms to three digits for $L\le10^{-2}$. This is the optimum for phase estimation. For absorption sensing there is no optimum to compare, by Sec.~\ref{sec:absorption}: the quantum Fisher information for the transmissivity is linear in the photon number for every state, N00N states included, so they share the flat exchange rate of the first row of Table~\ref{tab:classes}. Entangled probes therefore have a constant of their own for phase. It is $x_{\rm opt}=1$ with value $1/e$, in place of $x_{\rm opt}=1.5936$ with $0.6476$.

The sequential chain therefore beats the parallel entangled state at equal dose by
\begin{equation}
e\,h(x_{\rm opt}) = 1.7604 ,
\label{eq:seqvspar}
\end{equation}
which is the same factor that separates the dose bookkeeping of Eq.~(\ref{eq:dose}) from the naive one. That is not a coincidence. The naive bookkeeping is the parallel accounting applied to a sequential scheme.

The two cases are the ends of one family. Let $N$ entangled photons each make $m$ passes. The coherence now requires every one of the $N$ photons to survive all $m$ of its passes, so the payoff is $(Nm)^{2}q^{Nm}$, while the dose is $N$ times the survival-weighted dose of a single photon. Maximising over $m$ at fixed $N$, and then over $N$, returns $N=1$. In units of $1/L$ and in the limit $L\to0$ the optimised value falls monotonically with $N$: it is $0.648$ at $N=1$, $0.482$ at $N=2$, $0.439$ at $N=3$, and tends to $1/e=0.368$ as $N$ grows. A single recycled photon is optimal over the whole family. Throughout we follow the convention of Ref.~\cite{Yu2026}, with the loss in the sensing arm and the balanced readout taken in its unbalanced small-weight limit. Doubling the loss exponent, as balanced loss in both arms would, rescales the constants but leaves the ordering intact: we checked numerically that a single recycled photon remains the global optimum.

Entanglement therefore does not escape the constant of Eq.~(\ref{eq:h}). It loses to it. The reason is visible in the exponents. The $N$-fold phase of the entangled state is fully credited, since the payoff scales as $(Nm)^{2}$ and not as $N m^{2}$; but the same entanglement requires every photon to survive every pass, so the coherence dies as $q^{Nm}$ while the dose still grows as $N$. The phase enhancement and the fragility are one property, and per absorbed photon the fragility wins. The accounting is generous to the entangled state in one further respect. Its payoff sits in an $N$-photon observable, so reaching it requires number-resolving joint detection --- or, in lithographic application, an $N$-photon absorber --- which is granted here at unit efficiency \cite{Wildfeuer2026companion}; the recycled photon needs one click detector. Note also that the figure of merit is intensive: $N$ unentangled photons, each folded $m_{\rm opt}$ times through the sample in its own loop, attain the same $h(x_{\rm opt})/L$ per absorbed photon at $N$ times the rate. The statement is that entanglement across photons does not help, not that the experiment must wait for one photon. Note that this is a statement about the damage currency alone. Per photon \emph{sent}, or per channel use, entangled probes remain advantageous in the standard way, and nothing here contradicts that literature \cite{Lee2009}.

One case sits between the rows of Table~\ref{tab:classes}. The entangled Fock states $|m{::}m'\rangle$ of Ref.~\cite{Huver2008} trade phase for robustness: they acquire $(m-m')\varphi$ per pass rather than $N\varphi$, but a single lost photon does not carry full which-path information, so their coherence decays more slowly than $q^{Nm}$. They were introduced precisely because N00N states are too fragile under loss. Whether any $|m{::}m'\rangle$ probe beats a single recycled photon per absorbed photon is not settled by the family above, because its decoherence is not a simple power of the survival. We computed it, with the single photon priced in the same framework. The lossy state is block diagonal in the number of lost photons, each block is at most two dimensional, and the quantum Fisher information reduces to a single sum,
\begin{equation}
F = 2p^{2}(m-m')^{2}\sum_{k=0}^{m'}\frac{c_{m,k}^{2}\,c_{m',k}^{2}}{c_{m,k}^{2}+c_{m',k}^{2}},
\label{eq:mmqfi}
\end{equation}
with $c_{n,k}^{2}=\binom{n}{k}T^{\,n-k}(1-T)^{k}$, $T=q^{p}$ the transmission after $p$ passes, and dose $\tfrac12(m+m')(1-T)$; exact diagonalisation of the lossy two-mode state confirms Eq.~(\ref{eq:mmqfi}) to machine precision. Note that the payoff here is the quantum Fisher information of the lossy state itself, for probe and benchmark alike, not the detected-photon payoff of hypothesis (H1); it is the most generous accounting available to the entangled states, and they still lose. No state reaches the benchmark. The states of Eq.~(\ref{eq:mmqfi}) are symmetric superpositions, and the like-for-like benchmark is the single photon in the balanced interferometer, $w=\tfrac12$. Against it the two-photon N00N state attains $0.707$ of the single-photon value, the best state with $m\le5$ is $|5{::}1\rangle$ at $0.720$, and the envelope over all $(m,m')$ rises slowly with photon number: the optimal offset $m-m'$ grows as the square root of $m'$, and a Gaussian limit of the sum in Eq.~(\ref{eq:mmqfi}) fixes the supremum at $0.868$ as $m'\to\infty$. Against the single photon at its optimal readout of Sec.~\ref{sec:phase}, which is $1/0.853$ times better, the same three numbers read $0.603$, $0.614$ and $0.740$. The supremum is approached, never attained. The near-diagonal states are the weakest of all, because their dose grows with $m+m'$ while their phase carries only $m-m'$. The robustness that Ref.~\cite{Huver2008} bought per photon does not survive the change of currency. Two classes remain outside the comparison. Unbalanced $|m{::}m'\rangle$ superpositions, and arbitrary superpositions of the kind that optimise lossy phase estimation per photon \cite{Dorner2009}, have not been priced per absorbed photon. And Gaussian probes have: Refs.~\cite{Koppell2022,KoppellKasevich2022} show that a squeezed Gaussian probe reaches, for strong squeezing, the per-dose quantum limit of which the recycled photon attains $0.65$, so the recycled photon is optimal within the number-state family priced here, not among all probes.

One practical consequence is that the position of the optimum reports how the resource was spent. An optimum at one unit of probe uses times loss indicates a parallel arrangement, and an optimum at $1.59$ indicates a sequential one. The unit is passes times loss for a chain and photons times loss for a parallel state.

Koppell and Kasevich \cite{KoppellKasevich2022} and, independently, Ref.~\cite{Yu2026} obtain the parallel result for N00N states, quoting $N_{\rm opt}\simeq-1/\ln\eta$ and an optimality coefficient carrying $1/e$. Equation~(\ref{eq:parallel}) is that statement in the present notation. Entanglement thus enters the table as a second class, not as an exception.

\section{Detecting a faint companion below the Rayleigh limit}
\label{sec:subrayleigh}

Is there a faint companion next to a known source, closer than the Rayleigh limit? This is the detection variant of the two-point resolution problem, and it is a null experiment in the sense made precise next. The ideal exponents and the crosstalk ceiling are known \cite{LuKroviTsang2018,Schlichtholz2024}. What we add is the pricing per absorbed photon.

\subsection{Null experiments}
\label{sec:null}

A null experiment is an apparatus tuned so that one hypothesis forbids an outcome outright, the zero being a property of the unperturbed apparatus rather than of the object under test. The Zeno interrogator of Ref.~\cite{Wildfeuer2026} is one realisation; the dark odd mode of a centred spatial-mode sorter, treated below, is another. The following lemma, for a binary test of $H_0$ against $H_1$, is elementary. We state it in full because both configurations below and the discrimination row of Table~\ref{tab:classes} rest on its equality case.

\begin{Lemma}
Let a trial have outcome set $\Omega$, let $F=\{\omega: P(\omega|H_0)=0\}$ be the $H_0$-forbidden set, and let $p_F=P(F\,|\,H_1)>0$. Then the one-sided test ``decide $H_1$ if an outcome in $F$ has occurred at least once'' has type-I error zero and type-II error exactly $(1-p_F)^n$ after $n$ independent trials. The Chernoff information per trial \cite{Chernoff1952,CoverThomas2006,Nielsen2022} obeys
\begin{equation}
\begin{split}
C \;&=\; -\ln\!\Big(\min_{0\le s\le1}\sum_{\omega\notin F} P(\omega|H_0)^{1-s}P(\omega|H_1)^{s}\Big)\\
\;&\ge\; -\ln(1-p_F),
\end{split}
\label{eq:lemma}
\end{equation}
because the sum at $s=1$ equals $1-p_F$ and the minimum can only be smaller. The sum is a convex function of $s$, so equality holds if and only if its derivative at $s=1$ is non-positive, that is
\begin{equation}
D(Q\,\|\,P_0)\;\le\;-\ln(1-p_F),
\label{eq:lemmacond}
\end{equation}
where $Q(\omega)=P(\omega|H_1)/(1-p_F)$ is the $H_1$ distribution conditioned on the allowed set $\Omega\setminus F$, $P_0=P(\cdot|H_0)$, and $D$ is the relative entropy. Proportionality outside $F$, $Q=P_0$, is the special case $D=0$ and is not necessary.
\end{Lemma}
The one-sided test is therefore optimal whenever the hypotheses are so alike outside $F$ that the forbidden set carries the evidence, and in particular whenever $H_0$ has support on a single outcome, as for the Zeno chain and for the binary odd-mode reduction below, where $Q=P_0$ trivially. In general the one-sided test is optimal among tests that use only occurrences in $F$, and discrimination outside $F$ can only add to $C$.

The error is one-sided: $H_0$ is never falsely rejected. When both hypotheses have full support every likelihood ratio is finite, and no finite number of trials yields certainty. The divide between proof and statistics drawn in Ref.~\cite{Wildfeuer2026} is thus a consequence of support alone, and no optical arrangement is needed to produce it.

\subsection{Two configurations, one lemma}
\label{sec:twoconf}

Take an imaging system with a Gaussian point-spread function of width $\sigma$ and Hermite--Gauss spatial-mode demultiplexing \cite{TsangNair2016}. A point source at offset $s$ populates the first odd mode with probability
\begin{equation}
P(q{=}1\,|\,s) = \frac{s^2}{4\sigma^2}\,e^{-s^2/4\sigma^2},
\label{eq:hg01}
\end{equation}
so a single source on the sorter axis leaves that mode exactly dark. The lemma applies, with $p_1$ the odd-mode rate under $H_1$, and one odd-mode photon is conclusive.

Configuration (a) preserves the centroid. Under $H_0$ there is one source at the origin, under $H_1$ two half-intensity sources at $\pm d/2$ \cite{LuKroviTsang2018}. Then $p_1=d^2/16\sigma^2$, while direct imaging sees no first-order change in the profile,
\begin{equation}
C_{\mathrm{direct}}^{\mathrm{(a)}} \simeq \frac{d^4}{256\,\sigma^4},
\qquad
\frac{C_{\mathrm{null}}}{C_{\mathrm{direct}}} \simeq \frac{16\,\sigma^2}{d^2},
\label{eq:sym}
\end{equation}
which is unbounded as $d\to0$. Both constants agree exactly with Table~I of Ref.~\cite{LuKroviTsang2018}.

Configuration (b) is the exoplanet case: a known primary with a faint companion \cite{HuangLupo2021,GraceGuha2022,Deshler2025}. With relative intensity $\epsilon_r$ at offset $d$, and $\epsilon_r'=\epsilon_r/(1+\epsilon_r)$, the null rate is $p_1\simeq\epsilon_r'(d/2\sigma)^2$. Here the profile shifts at first order in $d$, so direct imaging is quadratic rather than quartic,
\begin{equation}
C_{\mathrm{direct}}^{\mathrm{(b)}} \simeq \frac{\epsilon_r'^{\,2} d^2}{8\,\sigma^2},
\qquad
\frac{C_{\mathrm{null}}}{C_{\mathrm{direct}}} \simeq \frac{2}{\epsilon_r'},
\label{eq:asym}
\end{equation}
bounded and independent of separation. The advantage is large for a faint companion --- $2\times10^{4}$ at $\epsilon_r=10^{-4}$ --- but the brightness contrast caps it. We verified both constants against exact numerical Chernoff exponents, with agreement better than $1\%$ for $d\le0.1\sigma$. Reference \cite{Arbitrary2026} obtains general quantum Chernoff exponents for arbitrary intensity distributions, but treats symmetric and commuting configurations; Eq.~(\ref{eq:asym}) is not contained there. The practical reading is that the unbounded win belongs to the symmetric splitting problem. Discovery next to a known star or fluorophore wins instead by the brightness contrast.

\subsection{The crosstalk ceiling}

Real sorters leak a fraction $\delta$ of centred-source light into the odd channel. Schlichtholz \emph{et al.}\ established the consequences \cite{Schlichtholz2024}. Their expansion of the crosstalk-affected Chernoff exponent contains both asymptotic branches, a quadratic collapse below the crosstalk scale and a logarithmically slow approach to the ideal scaling above it, together with the crossover between them. They further showed that separation-independent tests are destroyed by arbitrarily weak crosstalk, and constructed a semi-separation-independent replacement. Background noise has been treated numerically and experimentally \cite{Liu2026noise}, and misalignment in Refs.~\cite{DeAlmeida2021,Singular2026}.

\begin{figure}[H]
\includegraphics[width=0.6\textwidth]{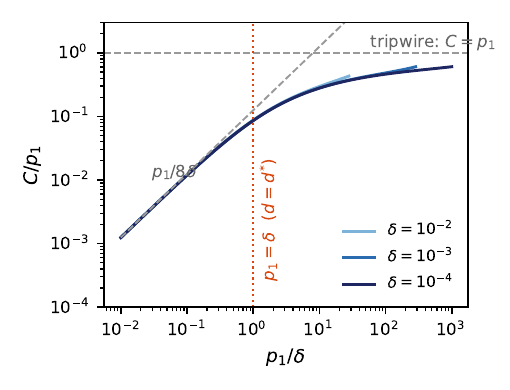}
\caption{\label{fig:ceiling}The crosstalk ceiling. Exact two-outcome Chernoff exponent $C$, normalised by the ideal value $p_1$, against $p_1/\delta$ for three crosstalk levels. Dashed lines: the asymptotes of Eq.~(\ref{eq:ceiling}), first obtained for the symmetric configuration in Ref.~\cite{Schlichtholz2024}. Dotted line: the crossover $p_1=\delta$, that is $d=d_{c}$ of Eq.~(\ref{eq:dstar}).}
\end{figure}

What this subsection adds is modest. The two-outcome reduction makes those asymptotics one-line consequences of Eq.~(\ref{eq:lemma}),
\begin{equation}
C \simeq
\begin{cases}
p_1, & p_1 \gg \delta,\\[2pt]
p_1^2/8\delta, & p_1 \ll \delta,
\end{cases}
\label{eq:ceiling}
\end{equation}
which collapse onto a single curve in $p_1/\delta$ as Fig.~\ref{fig:ceiling} shows. Transcribed to configuration (b), the crossover becomes a separation,
\begin{equation}
d_{c} \simeq 2\sigma\,\sqrt{\delta/\epsilon_r}
\label{eq:dstar}
\end{equation}
for $\epsilon_r\ll1$, and with $\epsilon_r\to\epsilon_r'$ in general. Below $d_{c}$ discovery is no longer free and the advantage per absorbed photon dies, just as the Zeno advantage dies for $N>N_{\rm opt}$.

\subsection{Dose accounting for active illumination}
\label{sec:dose}

Let the scene be actively illuminated, so that the primary and the possible companion absorb photons at rates $A_p$ and $A_c$, re-emit with quantum yield $Y$, and the emission is collected with efficiency $\eta_{\rm col}$. All of this is identical for the two measurement schemes, which differ only after the light has left the specimen. Three statements follow, in increasing order of content.

First, ratios are inherited. Any two measurements sharing the illumination and collection path have identical exponent ratios in every damage measure, because the conversion from collected to absorbed photons is a measurement-independent factor. The comparisons of Sec.~\ref{sec:twoconf} therefore carry over unchanged.

Second, if the whole specimen is fragile, exponents per absorbed photon are the per-collected-photon exponents multiplied by $\kappa=Y\eta_{\rm col}$. Nothing else changes.

Third, the fragile entity may be the companion itself. Is there a delicate structure beside a robust and known one? Normalise by the photons the companion absorbs. The odd-mode click rate is the total collected rate times $p_1$,
\begin{equation}
\eta_{\rm col}Y\,(A_p+A_c)\times\frac{A_c}{A_p+A_c}\Big(\frac{d}{2\sigma}\Big)^{2}
=\eta_{\rm col}Y A_c \Big(\frac{d}{2\sigma}\Big)^{2},
\label{eq:clickrate}
\end{equation}
so that per photon the companion absorbs
\begin{equation}
\frac{C_{\mathrm{null}}}{\nabs^{(c)}} = \eta_{\rm col}Y\,\Big(\frac{d}{2\sigma}\Big)^{2}.
\label{eq:companiontax}
\end{equation}
This is independent of the brightness contrast $\epsilon_r$, because the primary's brightness cancels between the odd-mode rate and the companion's share of the absorption. Faintness does not change what discovery costs the companion. Note that the invariance survives unequal emission properties. With distinct yields $Y_c$ and $Y_p$, both the odd-mode rate and the companion's dose scale with the companion alone, giving $\eta_{\rm col}Y_c(d/2\sigma)^{2}$, which is free of every property of the primary.

Direct imaging delivers $\eta_{\rm col}Y\epsilon_r' d^{2}/8\sigma^{2}$ in the same measure. That vanishes linearly as the companion grows fainter, so the damage it suffers per unit of certainty grows as $1/\epsilon_r$. The null mode charges the companion a fixed rate. Direct imaging charges it without bound.

Some representative numbers: at $\eta_{\rm col}Y\simeq0.1$ and $d=0.2\sigma$ one obtains about $10^{-3}$ conclusive answers per companion-absorbed photon, that is some $10^{3}$ absorbed photons per conclusive discovery. This sits comfortably inside the $10^{5}$ to $10^{6}$ photons a good organic fluorophore emits before it bleaches \cite{Lichtman2005,Demchenko2020}. Crosstalk changes this below the crossover. For $d<d_{c}$ the exponent is $p_1^{2}/8\delta$ by Eq.~(\ref{eq:ceiling}), and per photon the companion absorbs
\begin{equation}
\frac{C_{\mathrm{null}}}{\nabs^{(c)}} \simeq \eta_{\rm col}Y\,\frac{\epsilon_r'}{8\delta}\Big(\frac{d}{2\sigma}\Big)^{4},
\qquad d<d_{c},
\label{eq:companiontaxbelow}
\end{equation}
which vanishes linearly with the companion's brightness, exactly as direct imaging does: the $1/\epsilon_r$ divergence of the cost returns. The invariance of Eq.~(\ref{eq:companiontax}) is therefore a property of the crosstalk-free regime $d>d_{c}$. Since $d_{c}$ grows as $\epsilon_r^{-1/2}$, a sufficiently faint companion is always below it at a given sorter extinction, and Sec.~\ref{sec:outlook} shows that with current sorters this is the case for $\epsilon_r\lesssim10^{-2}$ at all sub-Rayleigh separations. In that regime the null mode still beats direct imaging, by the separation-dependent factor $d^{2}/16\sigma^{2}\delta$, but it no longer spares the companion.

\section{Relation to prior work}
\label{sec:prior}

The provenance of the results must be exact. The Zeno-chain optimum, Eq.~(\ref{eq:zeno}), is Ref.~\cite{Wildfeuer2026}, derived in Appendix~\ref{app:zeno}. The multi-pass optimum was found by Koppell and Kasevich, with a survival-weighted dose, together with the fraction $0.65$ of the per-dose quantum limit, the N00N value $1/e$, and the observation that chain interferometers with pass-dependent couplings approach the limit and squeezed Gaussian probes saturate it \cite{Koppell2022,KoppellKasevich2022}; and independently by Yu and co-workers for optical interferometry, as a single-loss-channel result in the photons-entering measure \cite{Yu2026}. What is ours is the statement of the three hypotheses as the structure these results share, the exact form of Eq.~(\ref{eq:prop}), the separation of $\alpha$ from $\epsilon$, the sequential-versus-parallel diagnosis of Sec.~\ref{sec:entangled} with the $|m{::}m'\rangle$ supremum of Eq.~(\ref{eq:mmqfi}), and the invariance of Eq.~(\ref{eq:companiontax}) together with its boundary, Eq.~(\ref{eq:companiontaxbelow}). The constants are not ours and are credited above.

For the detection problem the ideal exponents are due to Lu \emph{et al.}\ \cite{LuKroviTsang2018} for the symmetric pair, their Table~I agreeing exactly with Eq.~(\ref{eq:sym}), and to the exoplanet literature \cite{HuangLupo2021,GraceGuha2022,Deshler2025} for the faint companion, with general arbitrary-intensity exponents in Ref.~\cite{Arbitrary2026}. The crosstalk asymptotics and the crossover are due to Schlichtholz \emph{et al.}\ \cite{Schlichtholz2024}. Misalignment and background noise are treated in Refs.~\cite{DeAlmeida2021,Singular2026,Liu2026noise}, and crosstalk in separation estimation restores a Rayleigh-type curse \cite{Linowski2023}. Against this body Sec.~\ref{sec:subrayleigh} claims the framing through the lemma, the transcription of Eq.~(\ref{eq:dstar}), and the pricing of Sec.~\ref{sec:dose}, in particular the brightness-independent cost of Eq.~(\ref{eq:companiontax}) and its diverging direct-imaging counterpart. Multi-pass microscopy \cite{Juffmann2016} is the experimental precedent for $m$-fold dose advantages, and sub-Rayleigh discrimination has been demonstrated in Ref.~\cite{Zanforlin2022}. We verified the delimitations stated here against the full texts of Refs.~\cite{Yu2026,Arbitrary2026,Schlichtholz2024,LuKroviTsang2018,Liu2026noise}. Reference \cite{Yu2026} treats a single loss channel with dose defined as photons entering the sample channel, and Ref.~\cite{Arbitrary2026} does not contain the asymmetric known-primary form.

\section{Experimental outlook}
\label{sec:outlook}

For the phase problem the requirement is a low-loss recirculation loop: at a per-pass loss $\epsilon=10^{-2}$, which a modest cavity reaches, the dose advantage is already $G_{\max}\approx59$ at $m_{\rm opt}\approx145$ passes for $\alpha=10^{-3}$, and loss at the $10^{-3}$ level pushes it beyond $300$. For calibration against experiment, multi-pass microscopy has demonstrated a variance reduction of $4.8\pm0.8$~dB at constant damage, a factor of three, with classical light and modest pass numbers \cite{Juffmann2016}. The gap between that demonstrated factor of three and a ceiling of order $10^{2}$ is the engineering headroom the present analysis quantifies.

For the detection problem the single number is the sorter crosstalk $\delta$: the measured mean crosstalk intensity of a multi-plane light-conversion sorter is $1.7\times10^{-3}$, or $-28$~dB \cite{Boucher2020}. In the most sensitive separation-estimation experiment to date the crosstalk was already negligible against detector noise \cite{Rouviere2024}, so the practical $\delta$ in Eq.~(\ref{eq:dstar}) may be set by dark counts, which enter as an equivalent uniform background \cite{Liu2026noise}. Dark counts rather than efficiency set the parity threshold in Ref.~\cite{Wildfeuer2026companion} for the same reason. At $\delta=1.7\times10^{-3}$ the crossover sits at $d_{c}=4\sigma\sqrt{\delta}=0.17\,\sigma$ for the symmetric pair and at $0.27\,\sigma$ for a companion of relative intensity $\epsilon_r=0.1$. At $\delta=10^{-4}$ these improve to $0.04\,\sigma$ and $0.07\,\sigma$. For a very faint companion the crossover leaves the sub-Rayleigh regime altogether: $d_{c}=0.83\,\sigma$ at $\epsilon_r=10^{-2}$ and $\delta=1.7\times10^{-3}$, at the edge of the small-$d$ expansion. With current sorters the discovery of so faint a companion is crosstalk-limited at essentially all sub-Rayleigh separations, and the sorter extinction --- not the photon budget --- is then the technology to improve.

\section{Conclusion}
\label{sec:conclusion}

One function, $h(x)=x^{2}/(e^{x}-1)$, and one root, $x_{\rm opt}=1.59362$, govern the loss-limited optimum of fixed uniform multi-pass sensing per absorbed photon wherever three conditions meet: a quadratic payoff, the same survival on every pass, and a dose weighted by survival. Multi-pass electron microscopy, dose-efficient phase interferometry, phase sensing of a weakly absorbing object, and the interaction-free interrogator are one optimisation seen four times. The $0.65$ of Ref.~\cite{KoppellKasevich2022}, the $0.648$ of Ref.~\cite{Yu2026} and the $0.2625$ of Ref.~\cite{Wildfeuer2026} are the same number, carrying different geometric prefactors; for the interrogator the identity is asymptotic: it holds to leading order in the per-cycle loss for the conclusive-detection rate, with a relative correction of $-1.97\,\epsilon$, and to two percent for the Chernoff information.

The entangled probes priced here do not escape the constant but lose to it, because their coherence requires every photon to survive every pass while the dose still grows with photon number. Optimising over photon number and pass number together returns a single recycled photon, and the optimised value falls from $0.648$ to $1/e$ as the arrangement moves from sequential to parallel. Where the optimum sits reports which of the two an experiment is doing. The statement is one about the class: Gaussian probes and chains tailored pass by pass lie outside it and can exceed its ceiling, by at most a factor of about $1.5$ \cite{KoppellKasevich2022}.

The boundaries are priced in the same measure. Absorption estimation gains nothing, for anyone. Discrimination against absence gains a logarithm, through the null-experiment structure. Sub-Rayleigh discovery is a null experiment too, free only above the separation $d_{c}=2\sigma\sqrt{\delta/\epsilon_r}$. Above it the null mode charges a fragile companion a rate that the companion's faintness cannot raise, where direct imaging's charge grows without bound; below it the charge grows without bound for both.

Measure phase and not absorption when the sample is fragile. Cycle $x_{\rm opt}/(\epsilon+\alpha)$ times and no more. Use the null mode to discover and direct estimation to measure. And detune until the sample absorbs no more than the optics lose, and no further. All four say the same thing. Among single photons and path-entangled number states, the gentlest measurement of a fragile sample is also the simplest one: a single photon, recycled.

\paragraph{Funding:} This research received no external funding.

\paragraph{Data Availability Statement:} The original contributions presented in this study are included in the article. Further inquiries can be directed to the corresponding author.

\paragraph{Use of Artificial Intelligence:} This work was developed with the assistance of Claude (Anthropic), used for literature search and reference retrieval, symbolic and numerical verification of the results. All analytical results were checked by independent numerical computation, and all references were verified against publisher records. The author has verified the derivations reported here and is solely responsible for the content.

\paragraph{Conflicts of Interest:} The author declares no conflicts of interest.

\appendix
\section{The Zeno chain rate}
\label{app:zeno}

The interrogator of Ref.~\cite{Kwiat1999} rotates the photon's polarisation by $\theta=\pi/2N$ on each of $N$ cycles. Without an object the rotations add to $\pi/2$ and the photon leaves through the ``absent'' port. With an object in the rotated arm, each cycle projects the photon back: it reaches the object with probability $\sin^{2}\theta$ and is absorbed, or it survives the cycle with probability $\cos^{2}\theta$ and then suffers the parasitic loss $\epsilon$, so that the survival per cycle is $u=(1-\epsilon)\cos^{2}\theta$. After $N$ cycles it leaves through the ``present'' port with probability $u^{N}$, an outcome the absent hypothesis forbids; this is the forbidden set of the lemma. The dose is the absorption probability per cycle, weighted by survival to that cycle,
\begin{equation}
\nabs=\sin^{2}\theta\sum_{k=0}^{N-1}u^{k}=\sin^{2}\theta\,\frac{1-u^{N}}{1-u},
\label{eq:zenodose}
\end{equation}
so the conclusive-detection rate per absorbed photon is
\begin{equation}
\mathcal{R}_{\epsilon}(N)=\frac{u^{N}(1-u)}{\sin^{2}(\pi/2N)\,(1-u^{N})}.
\label{eq:zenoexact}
\end{equation}
If the parasitic loss acts before the object within a cycle, $\sin^{2}\theta$ is replaced by $(1-\epsilon)\sin^{2}\theta$, a change of relative order $\epsilon$. Without loss, $u=\cos^{2}\theta$ and for large $N$ the rate tends to $1/(N\theta^{2})=4N/\pi^{2}$, the lossless rate of Sec.~\ref{sec:zenosec}. With loss, $\sin^{2}\theta=(\pi/2N)^{2}[1+\mathcal{O}(N^{-2})]$ and
\begin{equation}
\mathcal{R}_{\epsilon}(N)=\frac{4}{\pi^{2}}\,\frac{(1-u)\,N^{2}u^{N}}{1-u^{N}}\,\big[1+\mathcal{O}(N^{-2})\big],
\label{eq:zenoprop}
\end{equation}
which has the algebraic form of the proposition with $A/B=4/\pi^{2}$, $q=u$ and $m=N$. Numerically the two sides of Eq.~(\ref{eq:zenoprop}) agree to $3\times10^{-4}$ at $N=50$ and to $2\times10^{-5}$ at $N=200$, for $\epsilon$ between $10^{-3}$ and $10^{-2}$.

The chain nevertheless does not satisfy (H1) to (H3) in their exact form. The survival $u=(1-\epsilon)\cos^{2}(\pi/2N)$ and the absorption probability $\sin^{2}(\pi/2N)$ both depend on the cycle number, whereas the proposition takes $q$ and $B$ independent of $m$. The connection to the parasitic loss $\epsilon$ alone is asymptotic. Hold $x=N\epsilon$ fixed and let $N\to\infty$. Then $-N\ln u=x+(\pi^{2}/4+x^{2}/2)/N+\mathcal{O}(N^{-2})$ and $(1-u)/\sin^{2}\theta=(4Nx/\pi^{2})\,[1+\pi^{2}/(4Nx)+\mathcal{O}(N^{-2})]$, so that
\begin{equation}
\mathcal{R}_{\epsilon}(N)=\frac{4}{\pi^{2}}\,\frac{h(N\epsilon)}{\epsilon}\,
\Big[1+\frac{c(x)}{N}+\mathcal{O}(N^{-2})\Big],\qquad
c(x)=\frac{\pi^{2}}{4x}-\frac{e^{x}}{e^{x}-1}\Big(\frac{\pi^{2}}{4}+\frac{x^{2}}{2}\Big).
\label{eq:zenoasym}
\end{equation}
The correction is of relative order $1/N$, that is of order $\epsilon$ at fixed $x$, and not $N^{-2}$. At $x_{\rm opt}$, where $e^{x}/(e^{x}-1)=2/x$, it equals $-(1+\pi^{2}/4x_{\rm opt}^{2})\,\epsilon=-1.97\,\epsilon$. The optimum therefore sits at $N_{\rm opt}=x_{\rm opt}/\epsilon+\mathcal{O}(1)$ with $\mathcal{R}_{\max}=(0.2625/\epsilon)\,[1-1.97\,\epsilon+\mathcal{O}(\epsilon^{2})]$, of which Eq.~(\ref{eq:zeno}) is the leading order. The coefficient belongs to the ordering of loss and object adopted here; with the loss acting before the object it is $-0.97$. The exact rate of Eq.~(\ref{eq:zenoexact}) confirms the expansion. At $\epsilon=10^{-2}$ its maximum over integer $N$ is $25.73$ at $N=160$, against $26.25$ at leading order; at $\epsilon=10^{-3}$ it is $261.95$ at $N=1594$, against $262.47$. The correction accounts for the difference in both cases. To the same leading order, by the lemma each conclusive trial carries $-\ln(1-u^{N})$ nats of Chernoff information; maximising $(4/\pi^{2})\,x^{2}[-\ln(1-e^{-x})]/[\epsilon(1-e^{-x})]$ instead of $(4/\pi^{2})h(x)/\epsilon$ gives $x=1.270$ and $0.300/\epsilon$, two percent above the information delivered at the rate optimum.

\end{document}